\documentclass[times,5p,twocolumn,final]{elsarticle}

\usepackage{lineno,hyperref}
\modulolinenumbers[5]

\usepackage{amssymb}
\usepackage{enumerate}
\usepackage{multirow}
\usepackage{amsmath}

\usepackage{array}
\usepackage{algorithmic}
\usepackage{booktabs}
\usepackage{underscore}

\journal{Journal of the Audio Engineering Society}

\begin{document}

\begin{frontmatter}

\title{ICSD: An Open-source Dataset for Infant Cry and Snoring Detection}

\author[1]{Qingyu Liu}
\author[1]{Longfei Song}
\author[2]{Dongxing Xu}
\author[1]{Yanhua Long\corref{cor1}}
\cortext[cor1]{The first two authors (Qingyu Liu, Longfei Song) contributed equally
to this work. Yanhua Long is the corresponding author. e-mail: yanhua@shnu.edu.cn}

\address[1]{Shanghai Engineering Research Center of Intelligent Education and Bigdata, Shanghai Normal University, Shanghai, China}
\address[2]{Unisound AI Technology Co., Ltd., Beijing, China}

\begin{abstract}

The detection and analysis of infant cry and snoring events are crucial
tasks within the field of audio signal processing. While existing datasets
for general sound event detection are plentiful, they often fall short in
providing sufficient, strongly labeled data specific to infant cries and
snoring. To provide a benchmark dataset and thus foster the research of
infant cry and snoring detection, this paper introduces the \textbf{I}nfant
\textbf{C}ry and \textbf{S}noring \textbf{D}etection (ICSD) dataset, a novel,
publicly available dataset specially designed for ICSD tasks.
The ICSD comprises three types of subsets: a real strongly labeled subset
with event-based labels annotated manually, a weakly labeled subset with only
clip-level event annotations, and a synthetic subset generated and labeled with strong
annotations. This paper provides a detailed description of the ICSD creation process,
including the challenges encountered and the solutions adopted.
We offer a comprehensive characterization of the dataset, discussing its limitations
and key factors for ICSD usage. Additionally, we conduct extensive
experiments on the ICSD dataset to establish baseline systems and offer insights
into the main factors when using this dataset for ICSD research.
Our goal is to develop a dataset that will be widely adopted by the community as a new open
benchmark for future ICSD research.

\end{abstract}


\end{frontmatter}


\section{Introduction}
\label{sec:intro}

In the field of machine perception, datasets are of critical importance. Particularly
in the domain of audio, each research direction has its specialized dataset, such as
the LibriSpeech dataset \cite{Librispeech} for Automatic Speech Recognition (ASR), the
VoxBlink dataset \cite{Voxblink} for speaker verification, the VoiceBank+DEMAND dataset \cite{VoiceBankDEMAND}
for speech enhancement, and the LJSpeech dataset \cite{LJSpeech} for speech synthesis, etc.
In the realm of sound event detection, the Detection and Classification of Acoustic Scenes
and Events (DCASE) has initiated and provided specialized datasets, such as the DESED dataset
for DCASE task4 \cite{DESED}. This dataset is targeted for applications in domestic environments,
focusing on 10 classes of sound events such as doorbell rings, dog barks, speech, and so on.
However, for application scenarios involving the detection of infant crying and snoring sounds,
there are currently no publicly available datasets in the literature.

The detection of infant crying and snoring has significant implications for both health and well-being.
Infant crying is a critical form of communication for infant, indicating various needs such as hunger,
discomfort, or distress. Continuous monitoring and prompt response to infant crying are essential,
particularly during the night when parents are sleeping. On the other hand, snoring is a common
symptom of sleep disorders such as sleep apnea, which can lead to serious health complications
if left untreated. Regular detection and analysis of snoring patterns can provide valuable
insights for medical diagnosis and treatment. In a domestic environment,
the development of effective detection systems for these sounds can greatly enhance the quality
of life for families, improve infant care, and contribute to sleep disorder diagnosis and treatment.

Numerous existing datasets have collected audio recordings of infant cry and snoring for potential acoustic event detection detection or classification studies. For instance, Google Research extracted 120,459 clips from the original Audioset\cite{Audioset} in 2020 for strong label annotation, including 556 infant cry audio files and 465 snoring audio files. However, the quantity of these strongly labeled data is not sufficient for sound event detection specific to infant crying and snoring. ESC-50 \cite{ESC50} comprises 40 five-second audio clips for each of 50 different classes, including categories for infant crying and snoring. However, rather than for detection tasks, this dataset has been specifically designed for multi-label sound event classification tasks. Baby Chillanto database \cite{BabyChillanto} includes 2267 one-second audio clips of infant crying. However, this dataset is primarily designed for direct medical pathologies classification and identification rather than routine monitoring. Two snoring datasets from Kaggle\footnote{https://www.kaggle.com/}, one for general snoring \cite{KaggleS} and another for female and male snoring \cite{KaggleFM}, provide 1500 one-second snoring audio clips totally. However, the majority of the samples in these two datasets are set against quiet backgrounds, which does not reflect the complexity of real-world scenarios.

\begin{table*}[!htbp]
  \caption{Basic Description of Source Datasets Used for ICSD Dataset Creation}
  \label{SourceDatasets}
  \vspace{0.2cm}
  \centering
  \scalebox{0.85}{
  \begin{tabular}{l|>{\centering\arraybackslash}p{4.8cm}|>{\centering\arraybackslash}p{4.8cm}|c}
    \toprule
    \textbf{Database} & \textbf{Creator} & \textbf{Data} & \textbf{Papers} \\
    \hline
    Audioset & Google Research & \textbf{Total 3911} (Infant cry 1815, Snoring 2096) & \cite{Audioset1, Audioset2, Audioset3, Audioset4, Audioset5, Audioset6, Audioset7, Audioset8} \\
    \hline
    BCD & National Institute of Astrophysics and Optical Electronics, CONACYT Mexico & \textbf{Total 2268} (Infant cry) & \cite{DonateACry2, BCD1, BCD2, BCD3, BCD4, BCD5, BCD6, BCD7, BCD8, BCD9} \\
    \hline
    Donate A Cry & \url{https://github.com/gveres/donateacry-corpus} & \textbf{Total 457} (Infant cry) & \cite{DonateACry1, DonateACry2, DonateACry3, DonateACry4, DonateACry5} \\
    \hline
    Self-collected database & Collected by Tareq Khan & \textbf{Total 500} (Snoring) & / \\
    \hline
    FM Snoring Dataset & Self-recorded & \textbf{Total 1000} (Snoring) & / \\
    \hline
    ESC-50 & Warsaw University of Technology, Institute of Electronic Systems & \textbf{Total 80} (Infant cry 40, Snoring 40) & \cite{ESC1, Audioset5, Audioset6} \\
    \hline
    SINS & DCASE 2018 Challenge Task5 & \textbf{Total 72984} (Absence 18860, Cooking 5124, Dish washing 1424, Eating 2308, Other 2060, Social activity 4944, Vacuum cleaning 972, Watching TV 18648, Working 18644) & \cite{DESED} \\
    \hline
    MUSAN & Center for Language and Speech Processing, The Johns Hopkins University & \textbf{Total 10831} (Music 660, Noise 930, Speech 426) & \cite{MUSAN1} \\
    \bottomrule
  \end{tabular}}
\end{table*}

The existing datasets are either too small or not readily adaptable for research purposes in the joint sound event detection of infant crying and snoring sounds. These limitations of current available resources highlight the need for a large and easily accessible dataset to facilitate more advanced research in the field of infant cry and snoring detection. In this study,
we aim to create a new open-source dataset that can foster significant advancements in the field of infant cry and snoring detection. The main contributions are as follows:
\begin{enumerate}[1)]
\item We present the open-source Infant Cry and Snoring Detection (ICSD) dataset, a
unique resource that includes weakly labeled, synthetic, and real strongly labeled
data annotations.
\item We establish three baseline systems on the ICSD dataset, two of which are derived from the baselines of DCASE Task 4 Challenge \cite{DESED}, while the third is a novel approach proposed by our team. This setup provides a comprehensive reference point for future ICSD studies.

\item We provide a detailed analysis and discussion of the results and challenges in infant cry and snoring detection. By presenting the ICSD dataset and our preliminary findings, we hope to promote the future development in the detection of infant cry and snoring.
\end{enumerate}

The ICSD dataset and the baseline are publicly available at \url{https://github.com/QingyuLiu0521/ICSD/}.

\section{Source Datasets}

\begin{table*}[!htbp]
\caption{Overall Statistics of ICSD dataset (\#clips, 10s/clip).}
\label{tab:ovel}
\vspace{0.2cm}
\centering
\scalebox{0.97}{
\begin{tabular}{l|c|c|c|c|c|c}
\toprule
\multirow{2}{*}{\textbf{Set}} & \multicolumn{2}{c|}{\textbf{Train}} & \multicolumn{2}{c|}{\textbf{Validation}} & \multicolumn{2}{c}{\textbf{Test}} \\
\cline{2-7}
& \textbf{InfantCry} & \textbf{Snoring} & \textbf{InfantCry} & \textbf{Snoring} & \textbf{InfantCry} & \textbf{Snoring} \\
\hline
Weakly labeled & 1699 & 1577 & 189 & 176 & None & None \\
\hline
Real strongly labeled & 338 & 305 & 43 & 39 & 43 & 39 \\
\hline
Synthetic strongly labeled & 4000 & 4000 & 500 & 500 & 500 & 500 \\
\bottomrule
\end{tabular}}
\end{table*}

Table \ref{SourceDatasets} introduces the source datasets utilized in this research. These datasets covers a variety of audio samples, including the background sounds from various environments, the foreground snoring and infant cries.  Each dataset is unique in its composition, providing a comprehensive range of audio samples that significantly aid the data wide-domain coverage of our research. The table provides details about the creator of each dataset, the total number of data clips/recordings it contains, and references to papers that have previously utilized the dataset. This diversity of data have allowed us to develop and validate our models effectively. A brief description of each source dataset is presented below.

\textbf{Audioset}: AudioSet \cite{Audioset} is a comprehensive audio event dataset comprising over 2 million human-annotated, 10-second clips from YouTube. The dataset uses a hierarchical ontology of 632 event classes for annotation, allowing for complex labelling of sounds. The aim of AudioSet is to aid in developing high-performance audio event recognizers, similar to ImageNet's impact in the image domain. In 2020, additional strong labelling was performed on select clips, with annotators marking every distinct sound event and indicating their start and end times. AudioSet's strongly labeled data includes 456 classes. Given our research focus on Infant Cry and Snoring Detection, we only extracted the Infant Cry and Snoring categories from AudioSet, including
1713 Snoring and 1391 Infant Cry clips of weakly labeled data, and
383 Snoring and 424 Infant Cry clips of event-based strongly labelled data.

\textbf{Baby Chillanto Database (BCD)}: The Baby Chillanto Database \cite{BabyChillanto} is a publicly available resource compiled by the National Institute of Astrophysics and Optical Electronics, CONACYT, Mexico. It is specifically designed for infant cry pathology classification tasks, featuring a diverse array of infant crying sounds. Each infant cry audio was segmented into one-second duration, and is grouped into five categories, including asphyxia, deaf, hunger, normal and pain. In total, the BCD comprises 2,268 infant cry samples, making it a comprehensive resource for researchers studying infant cries.

\textbf{Donate A Cry}: Donate A Cry \cite{DonateACry} is a database designed to aid in identifying the needs of infants through their crying patterns. The cries in the dataset were recorded from babies aged between 0 and 2 years old. Following data cleaning, the dataset was streamlined to include five primary categories: hunger, burping, belly pain, discomfort, and tiredness. The dataset comprises a total of 457 files, with each audio clip having a duration of 7 seconds.

\textbf{Self-collected database}: This self-recorded snoring dataset \cite{KaggleS} is an organized collection of audio samples, all of which are one-second clips, divided into two primary classes: one for snoring sounds and the other for non-snoring sounds.
Each class has 500 samples and self-collected from various online sources.
Among the 500 snoring samples, 363 samples consist of snoring sounds of children, adult men and adult women without any background noise. The remaining 137 samples include snoring superimposed on non-snoring background sounds.

\textbf{FM Snoring Dataset}: The FM Snoring Dataset \cite{KaggleFM} is Female and Male Snoring Dataset is a well-structured collection of audio samples, each of which is a one-second clip with a sampling rate of 44,100 Hz. This dataset is categorized into two distinct classes, each containing 500 audio clips, representing snoring sounds from females or males.

\textbf{ESC-50}: ESC-50 \cite{ESC50} is a collection of 2000 environmental recordings evenly distributed across 50 classes of various common sound events. These are grouped into 5 categories: animal sounds, natural soundscapes, human sounds, interior/domestic sounds, and exterior/urban noises, with each category containing 10 classes. Each class contains 40 clips with 5 seconds long for each. In our research, we only extracted the portion of Infant Cry and Snoring data from the ESC-50 dataset.

\textbf{SINS}: SINS \cite{SINS} is a comprehensive collection of continuous recordings from a person spending a week in a holiday home. This dataset is meticulously organized with each file segmented into 60-second intervals, and the events categorized into specific activities such as absence, cooking, dish washing, eating, social activities, vacuum cleaning, watching TV, and working. For our research, we utilized a derivative of the SINS dataset, specifically used in the DCASE 2018 Challenge Task 5 \cite{DCASE2018}. This subset provided us with 72,984 clips of 10-second duration each, yielding approximately 200 hours of data.

\textbf{MUSAN}: MUSAN \cite{MUSAN} is a dataset that holds approximately 109 hours of audio comprising music, speech, and noise. It embraces diversity with music from various genres, speech from twelve different languages, and a broad range of noises.

\section{ICSD Dataset}
\label{sec:icsd}

Table \ref{tab:ovel} provides an overview of our proposed ICSD dataset. The entire dataset is divided
into three subsets: train, validation, and test sets for building the ICSD system. Three types of
data clips are collected to construct these subsets: weakly labeled clips with clip-level event
annotations, real strongly labeled clips with manually annotated event-based time-stamps, and
synthetic clips with synthesized event-based time-stamps. Since we focus solely on the ICSD event
detection problem rather than an audio tagging task, we did not include the weakly labeled data in the
test set.

As shown in Table \ref{tab:ovel}, we have achieved a relatively balanced sample distribution for the
infant cry and snoring detection system in each subset. For instance, we have 1,699 and 1,577 weakly
labeled clips of 10-second duration for the infant cry and snoring training sets, respectively. For
the real strongly labeled data, there are 338 clips for infant cry model training and 305 clips for
snoring model training. In this version of ICSD, we have created 8,000 clips for training and 1,000
clips for model validation and testing.

All the data clips in Table \ref{tab:ovel} are derived from the eight source datasets as
listed in Table \ref{SourceDatasets}. Due to the variations in audio formats, sampling rates,
recording channels, and audio quality among different sources, these datasets were not directly used
for ICSD data generation. Instead, a data format unifying process was conducted to
clean up all these source samples. The following sections provide detailed descriptions of the data
cleaning process and the characteristics of the weakly labeled, real strongly labeled, and synthetic
strongly labeled data.

\subsection{\textbf{Data Format Unifying}}
\label{subsec:sdf}

For all the eight source datasets in Table \ref{SourceDatasets}, we use
ffmpeg \cite{ffmpeg} to convert those multi-channel audio files into single-channel ones,
and down-sampling them into 16 KHz sampling rate.

\subsection{\textbf{Weakly Labeled Data Creation}}
\label{subsec:wldcr}

Based on the cleaned-up source data clips presented in Section \ref{subsec:sdf},
we first created the weakly labeled data (WLD) portion of the ICSD dataset.
The detailed statistics are summarized in Table \ref{tab:dsold}.
The Infant Cry category of our WLD is derived from three sources: Audioset, Donate A Cry, and ESC-50.
Meanwhile, the Snoring category is derived from two sources: Audioset and ESC-50.This results in a total of 1888 clips for Infant Cry and 1753 clips for Snoring. We divided the
samples within this part of the dataset into a 9:1 ratio for training (3276 samples) and validation
(365 samples). This diverse collection of WLD covers a wide range of sound variations, potentially
enhancing the ICSD system's generalization ability and effectiveness on unseen data.

\begin{table}[!htbp]
  \caption{Detail Statistics of Weakly Labeled Data (\#clip, 10s/clip).}
  \label{tab:dsold}
  \vspace{0.05cm}
  \centering
  \scalebox{0.97}{
  \begin{tabular}{l|l|c}
    \toprule
\textbf{Category} & \textbf{Data Source} & \textbf{\#Clips}  \\
    \hline
\multirow{3}{*}{Infant Cry} & Audioset  & 1391 \\
 & Donate A Cry  & 457 \\
 & ESC-50  & 40 \\
    \hline
\multirow{2}{*}{Snoring} & Audioset  & 1713 \\
 & ESC-50 & 40 \\
    \bottomrule
  \end{tabular}}
\end{table}

\subsection{\textbf{Real Strongly Labeled Data Creation}}
\label{subsec:rsldr}

The real strongly labeled data in our ICSD dataset is directly sourced
from Audioset, all these data clips are with human labeled target event-based
time-stamps. The detail statistics are shown in Table \ref{tab:srsld}.
We divide the clips within this part of dataset into a ratio of 8:1:1
for training (647 clips), validation (80 clips) and testing (80 clips).

\begin{table}[!ht]
\caption{Statistics of Real Strongly Labeled Data (\#clip, 10s/clip).}
\label{tab:srsld}
\vspace{0.05cm}
\centering
\scalebox{0.97}{
\begin{tabular}{l|c|c|c|c}
\toprule
\textbf{Category} & \textbf{Data Source} & \textbf{Train} & \textbf{Validation} & \textbf{Test} \\
\hline
Infant Cry & Audioset (424) & 340 & 42 & 42 \\
\hline
Snoring & Audioset (383) & 307 & 38 & 38 \\
\bottomrule
\end{tabular}}
\end{table}

\begin{figure}[!htbp]
\centering
\includegraphics[width=0.9\linewidth]{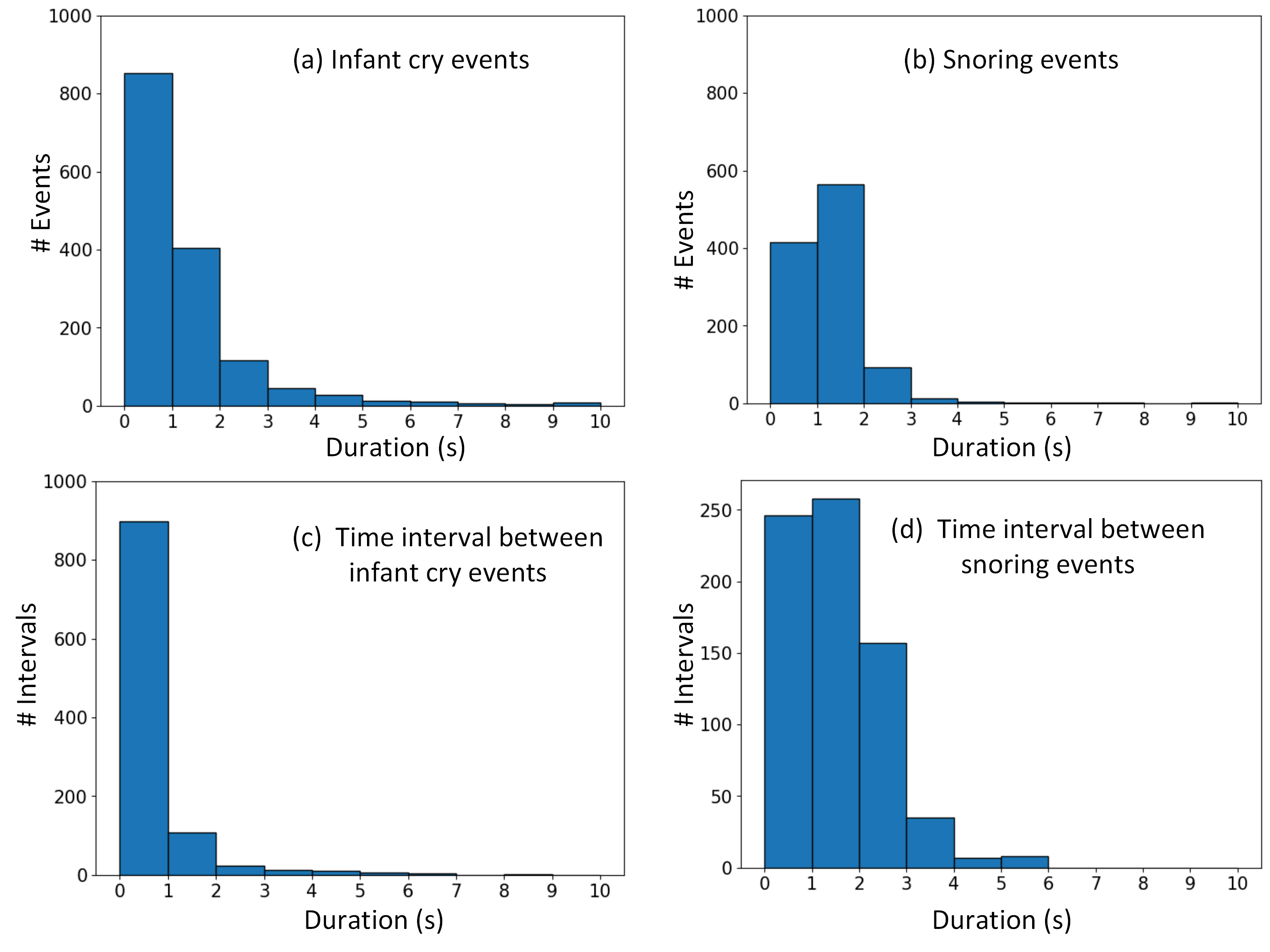}
\caption{Duration histogram of events and time intervals.}
\label{histogram}
\end{figure}

In addition, in Fig.\ref{histogram}, we plot the duration histogram of both
infant cry and snoring events as described in Table \ref{tab:srsld},
along with their time intervals between events. Fig.\ref{histogram}(a) and (b) show
the duration distribution of the target events, while Fig.\ref{histogram}(c) and (d)
illustrate the distribution of time intervals between two events in each recording clip.
These figures aid in understanding the temporal characteristics of the sound events,
which are fundamental for effective feature extraction and model training.

As illustrated in Fig.\ref{histogram}(a), the infant cry events mostly have
durations within the first two seconds. There is a notable peak at the outset,
indicating a high occurrence of very brief cries. The sharp decline in event
frequency as duration increases highlights the transient nature of infant cries.
These brief events imply the necessity of feature extraction techniques
tailored to capturing these rapid, intense bursts of sound effectively.
In Fig.\ref{histogram}(b), we see the snoring events tend to have slightly
longer durations, with the most common durations ranging from one to four seconds.
This suggests that snoring, while still brief, includes somewhat longer sounds
compared to infant cries, necessitating a distinct approach to feature analysis
capable of accommodating these long acoustic characteristics.

Examining the duration distribution of intervals between events,
Fig.\ref{histogram}(c) highlights that intervals of infant cry events
are primarily less than one second, indicating a high frequency of closely
spaced cry events. This pattern suggests the need for algorithms capable
of detecting rapid sequences of sound, which could signify specific behaviors or needs.
In contrast, the time interval distribution for Snoring events shown in
Fig.\ref{histogram}(d) demonstrates more extended periods between occurrences,
with intervals more evenly distributed up to three seconds. This regularity in event
spacing provides a clear rhythmic pattern that feature extraction or acoustic
modeling techniques must capture to effectively differentiate snoring from other sounds.

\subsection{\textbf{Synthetic Strongly Labeled Data Creation}}
\label{sec:syntheticdata}

\begin{figure*}[!htbp]
\centering
\includegraphics[width=0.75\linewidth]{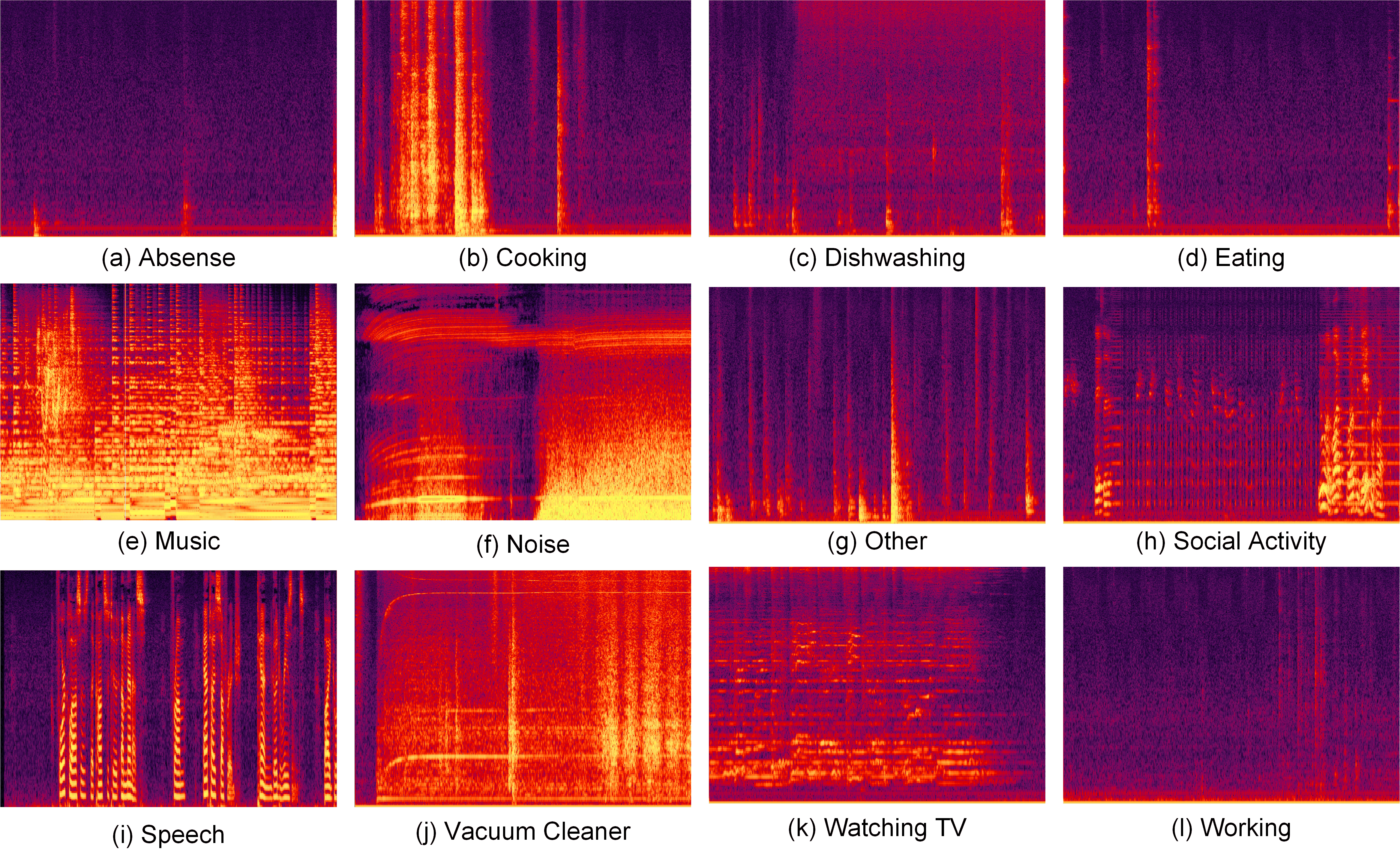}
\caption{Spectrogram samples of 12 types of background sounds.}
\label{backgroundspectrogram}
\end{figure*}

The synthetic strongly labeled (SSL) data in our ICSD dataset is specifically
generated to augment the existing data and provide a more robust training set
for the ICSD system training.

\subsubsection{\textbf{Source Material}}

The source material used to synthesize the SSL dataset can be divided into
two groups: the foreground events and background sounds. Details of these
sources are presented as bellow.

\textbf{Foreground Events}: In our study, foreground sounds are individual
sound events including both events segments of infant cry and snoring (they are
also called target events in ICSD system). As shown in Table \ref{tab:sfs},
there target events are all extracted from the source datasets as listed in
Table \ref{SourceDatasets}, specifically, the infanct cry events are extracted
from BCD and Audioset, while the snoring events are extracted from both the
self-collected database and FM Snoring dataset that directly downloaded from
the Kaggle website \footnote{https://www.kaggle.com/}.
To ensure the quality of synthetic data, we discarded foreground events with
poor quality including the asphyxia and deaf segments of BCD as well as any
foreground sounds with duration less than 250ms.
As in Table \ref{tab:sfs}, all these cleaned-up foreground events are then
divided into a ratio of 9:1 for the synthesize of
training and validation sets, and the synthesize of the test set.

\begin{table}[!htbp]
\caption{Statistics of Foreground Events.}
\label{tab:sfs}
\vspace{0.2cm}
\centering
\scalebox{0.97}{
\begin{tabular}{l|l|c|c}
\toprule
\textbf{Category} & \textbf{Source}(\#Events) & \textbf{Train \& Validation} & \textbf{Test} \\
\hline
\multirow{2}{*}{Infant Cry} & BCD (1048) & 943 & 105 \\
 & Audioset (1354) & 1211 & 143 \\
\hline
\multirow{2}{*}{Snoring} & Kaggle (1500) & 1350 & 150 \\
 & Audioset (1058) & 946 & 112 \\
\bottomrule
\end{tabular}}
\end{table}

\textbf{Background Sounds}: The background sounds used in our study consist of individual
ambient sound clips, each segmented into 10-second intervals. To avoid the dominance of
speech and music backgrounds, we did not use all the cleaned-up clips from the SINS and MUSAN
datasets (in Table \ref{tab:ovel})  for our ICSD backgrounds in SSL data synthesis. Instead,
we randomly extracted 15 hours of speech data and 10 hours of music data from the MUSAN
dataset and combined them with the SINS dataset as background sounds. Detailed information is
provided in Table \ref{tab:sbss}, which includes 11 different types of background
environment sounds and an additional category named `Other' for miscellaneous background
sounds. During SSL synthesizing, all the backgrounds in Table \ref{tab:sbss} are
divided into a ratio of 9:1 for the synthesis of training plus validation sets, 
and the synthesis of the test set. Fig. \ref{backgroundspectrogram} illustrates a randomly
picked sample background spectrogram for each category of background.

\begin{table}[!htbp]
\caption{Statistics of Background Sound Segments.}
\label{tab:sbss}
\vspace{0.2cm}
\centering
\scalebox{0.97}{
\begin{tabular}{l|c|c}
\toprule
\textbf{Category} & \# \textbf{Segments (10s)} & \textbf{\# hrs} \\
\hline
Absence & 18860 & 52.4 \\
Cooking & 5124 & 14.2 \\
Dish washing & 1424 & 4.0 \\
Eating & 2308 & 6.4 \\
Social activity & 4944 & 13.7 \\
Vacuum cleaning & 972 & 2.7 \\
Watching TV & 18648 & 51.8 \\
Working & 18644 & 51.8 \\
Speech & 5400 & 15 \\
Music & 3600 & 10 \\
Noise & 1831 & 5.1 \\
Other & 2060 & 5.7 \\
\hline
Total & 83815 & 232.8 \\
\bottomrule
\end{tabular}}
\end{table}

\subsubsection{\textbf{SSL Data Synthesizing Process}}
\label{subsubsec:ssldsp}

\begin{figure}[!htbp]
  \centering
  \includegraphics[width=0.9\linewidth]{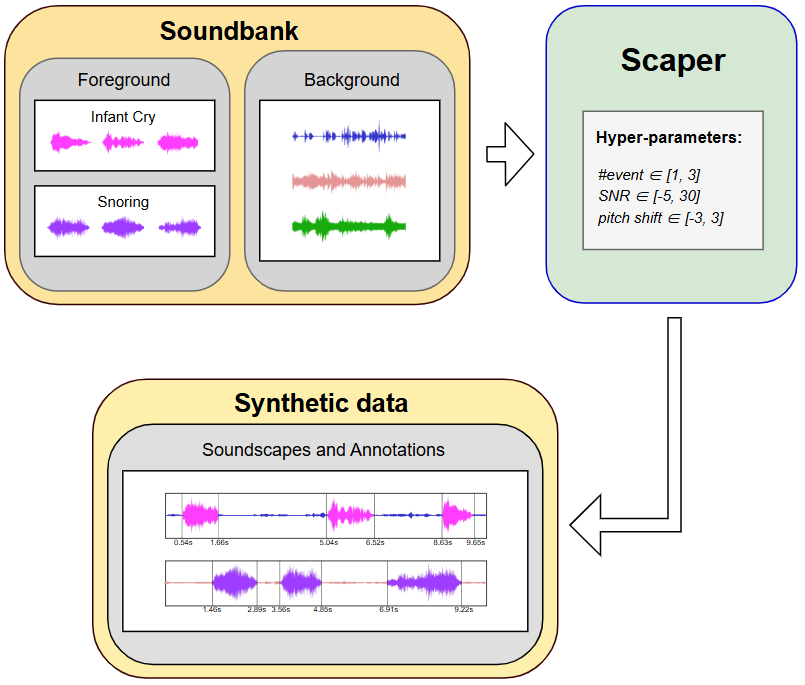}
  \caption{Illustration of SSL data synthesizing.}
  \label{synth}
\end{figure}

Given the above provided foreground events and background sound clips,
we synthesized the SSL subset of the ICSD dataset using the Scaper toolkit, as
illustrated in Fig. \ref{synth}. Scaper \cite{scaper} is a library specifically designed for the
synthesis of strongly labeled data and has been widely utilized in the DCASE
Challenges \cite{DESED}. Scaper can create a rich, customizable dataset from
collections of foreground and background sounds. It employs probabilistic
definitions to generate diverse audio scenes, allowing for the manipulation of
individual sounds with transformations such as pitch shifting and time stretching.
This toolkit is particularly valuable for sound event detection in diverse
environments, addressing the scarcity of strongly labeled audio data necessary for
training and evaluating machine listening systems.

The parameters for Scaper to generate the synthetic data SSL were carefully chosen to control the
characteristics of the soundscapes. These include the duration of each synthetic clip,
the minimum and maximum number of events in each clip, the level of reverberation, the foreground-background
signal-to-noise ratio (FB-SNR), and the range of pitch shifts applied to the events. The specific values of
these parameters are detailed in Table \ref{tab:pissdg}.

\begin{table}[!htbp]
\caption{Parameters in Scaper for SSL Data Generation.}
\label{tab:pissdg}
\vspace{0.2cm}
\centering
\scalebox{0.97}{
\begin{tabular}{l|c}
\toprule
\textbf{Parameter} & \textbf{Value} \\
\hline
Duration & 10s \\
Min event & 1 \\
Max event & 3 \\
Reverb & 0.1 \\
FB-SNR & (uniform, -5, 30) \\
Pitch shifts & (uniform, -3.0, 3.0) \\
\bottomrule
\end{tabular}}
\end{table}

Based on the above Scaper settings, we finally generated a total of 10,000 synthetic
clips for the current SSL dataset, as detailed in Table \ref{tab:ovel}.
This SSL dataset includes 4,000 training clips each for infant cry and snoring,
along with 500 clips each for validation and testing in both categories.

\subsection{\textbf{Ontology Data Release}}

\begin{figure}[t]
  \centering
  \includegraphics[width=0.8\linewidth]{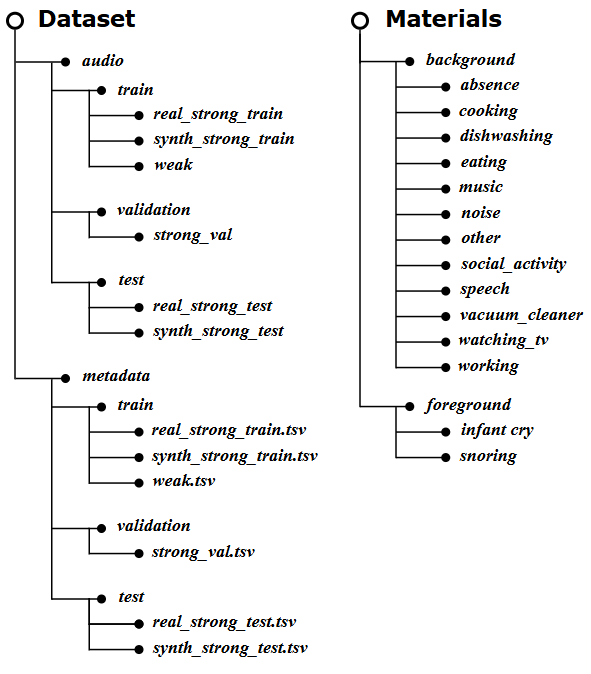}
  \caption{The structure of the ICSD dataset ontology.}
  \label{folder}
\end{figure}

 Fig.\ref{folder}
illustrates the organized structure where audio files are stored in the audio folder and
event time-stamp annotations in the metadata folder, each further categorized into train, validation,
and test subfolders. Furthermore, we have also released the source materials for generating synthetic strongly labeled data, including foreground events and background events. 

\section{\textbf{ICSD Systems}}
\label{sec:iscdsys}

We build three main ICSD systems on our ICSD dataset to serve as initial references for further research
on infant cry and snoring detection. These systems include an MT-CRNN system, a CRNN+BEATs system,
and an enhanced CRNN+BEATs system. The first two
architectures are adapted from the DCASE 2023 Challenge Task 4 baseline systems \cite{DCASE2023task4}, while the
competitive system is our improved version of CRNN+BEATs. Details of these three systems are presented
in the following sections.

\begin{figure*}[!htbp]
\centering
\includegraphics[width=0.9\textwidth]{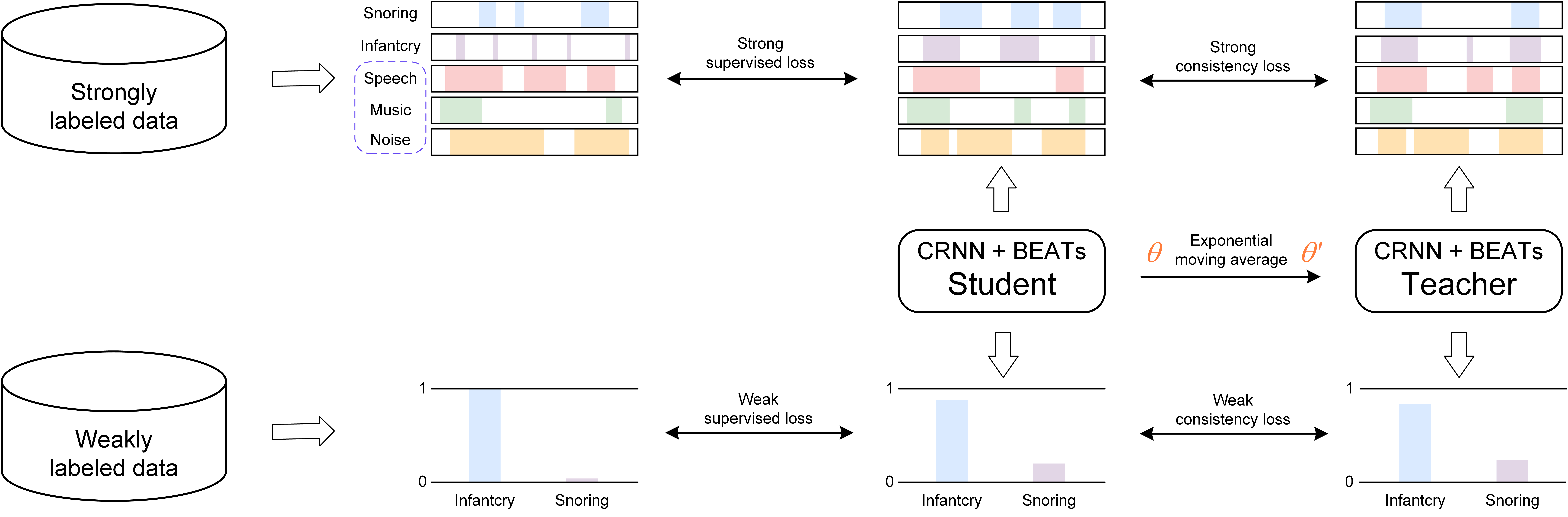}
\caption{Training procedure illustration of the Competitive CRNN+BEATs model.}
\label{competitivemodel}
\end{figure*}

\subsection{\textbf{MT-CRNN}}

The MT-CRNN ICSD system is based on the Mean-Teacher Convolutional Recurrent Neural Network \cite{MTCRNN},
it consists of a student model and a teacher model, both sharing same architecture
but serving different purposes. The model architecture is a fusion of a Convolutional
Neural Network (CNN) and a Recurrent Neural Network (RNN), topped with an attention layer.
The RNN output offers strong frame-level predictions, while the attention layer output
provides segment/clip-level predictions (weak predictions).

The student model is trained on a mix of strongly and weakly labeled data.
Binary cross entropy (BCE) is used to compute the supervised learning loss at the frame
level for strongly labeled data and at the clip level for weakly labeled data.
However, the teacher model is not directly trained. Instead, the teacher model's parameters
are updated using an Exponential Moving Average (EMA) approach\cite{EMA},
where the weights are a moving average of the student model's weights.
This procedure is defined as:
\begin{equation}
{\theta}'_{t} = \alpha \cdot {\theta}'_{t-1} + (1-\alpha)\cdot {\theta}_{t}
\end{equation}
where $\alpha$ is the decay hyperparameter used to control how far the EMA reaches into the
previous training history, ${\theta}'_{t}$ and ${\theta}_{t}$ are the current
model weights of the teacher and student model, respectively.

In this MT-CRNN, the teacher model contributes to the training of the student model via a
consistency loss (mean-squared error) for both strong and weak predictions. The
whole MT-CRNN model training is performed using four distinct losses combination
as defined in Eq.(\ref{eq:clc}),
\begin{equation}
\label{eq:clc}
\mathcal{L} = \mathcal{L}_{BCE}^{weak} + \mathcal{L}_{BCE}^{strong} +  \gamma \cdot \left( \mathcal{L}_{MSE}^{weak} + \mathcal{L}_{MSE}^{strong} \right)
\end{equation}
where $\gamma$ 
denotes the loss weight,
$\mathcal{L}_{BCE}^{\ast}$ and $\mathcal{L}_{MSE}^{\ast}$ are the
supervised BCE loss and teacher-student MSE consistency loss, respectively.
All the losses with `weak' subscript are computed using
the weakly labeled training data, while the losses with `strong' subscripts
are computed using the strongly labeled training data.
More details can be found in \cite{DESED}.

\subsection{\textbf{CRNN+BEATs}}

The CRNN+BEATs model builds upon the MT-CRNN model by incorporating the BEATs features.
BEATs \cite{Audioset5}, is an iterative audio pre-training framework to learn
Bidirectional Encoder representation from Audio Transformers. The BEATs
model is trained to extract high-level non-speech audio semantics
information using iterative self-supervised learning. In our study, the CRNN+BEATs
system architecture is the same as the one used in DCASE 2023 Task 4 Challenge baselines \cite{DCASE2023task4},
but we train it on our ICSD dataset for infant cry and snoring detection.

In the CRNN+BEATs model, the integration of BEATs feature representation and the CNN output
is performed by a linear transformation followed by a layer normalization.
The transformed feature map and CNN output are then concatenated and passed through
another linear transformation to match the dimension of the GRU input of the CRNN.
Our CRNN+BEATs model is trained in the same way as the MT-CRNN model. The inclusion of BEATs
feature representations provide additional complementary information to the CRNN latent
representations, which can potentially improve the performance of the model in
sound event detection tasks.

\subsection{\textbf{Competitive CRNN+BEATs}}

The Competitive CRNN+BEATs model is an improved version of the CRNN+BEATs model.
It incorporates three additional categories, speech, music and noise, into the real strongly labeled data.
This means we train a five-class SED system to perform the ICSD event detection,
rather than focusing solely on infant cry and snoring
detection. For each of speech and music categories, we downloaded 400 clips of strongly labeled data from
Audioset, dividing them into 350 clips for training and 50 clips for validation. For the noise category, we directly pick 140 samples from the background clips for model training. The training process for
the Competitive CRNN+BEATs model follows a similar procedure to that of the CRNN+BEATs model, as detailed in
Fig.\ref{competitivemodel}. In this case, the $\mathcal{L}_{*}^{strong}$
in Eq.(\ref{eq:clc}) are with all the five types of strongly labeled data (for the noise category, we labeled the whole clip as noise strong label in this case).

By incorporating additional categories for speech, music and noise, the model is better equipped to handle acoustic
confusion between speech, music, noise and target sound events, potentially improving its performance in ICSD
tasks. The motivation behind the proposed Competitive CRNN+BEATs model stems from our extensive preliminary
experiments, which showed that infant cry sounds have the most strong acoustic confusion with music, speech and noise, leading to high false positive rates during non-target environmental background sound testing.

\section{\textbf{Experiments and Results}}

\subsection{\textbf{Configurations}}

All our models were trained using the Adam optimizer. The batch size of the MT-CRNN system was set to 36, including 24 for strongly labeled data and 12 for weakly labeled data.Meanwhile, the batch size of the CRNN+BEATs system was set to 48, with 24 strongly labeled data and 24 weakly labeled data.
The learning rate was warmed up for 50 epochs, and
then decayed to 0.001. The models were trained over a total of 200 epochs.
The training loss weights $\gamma$
in Eq.(\ref{eq:clc}) are defined in Eq.(\ref{eq:sfactor}): 

\begin{equation}
\label{eq:sfactor}
\text{$\gamma$} =C \times e^{\:\beta \:\times \:\left(1 - \frac{\textit{s}}{\textit{N} \:\times\: \textit{L}}\right)^2}
\end{equation}
where $C$ = 2, represents the constant maximum value used for the self-supervised loss weight, \text{$\beta$} is the exponent parameter set to -5.0, \textit{s} is the current training step, \textit{N} is the number of warm-up epochs, set to 50. \textit{L} is the length of each epoch in terms of training steps.



\subsection{\textbf{Evaluation Metrics}}

To validate the performance of our models on the proposed ICSD, we use both
intersection-based F1 \cite{SED2, Intersection1} and segment-based F1 (Seg-F1) scores
\cite{SEGMENT1, SEGMENT2} as primary evaluation metrics. The intersection-based
F1 (Inter-F1) is calculated using a detection tolerance criterion (DTC) and ground truth
intersection criterion (GTC) set to 0.5, and a cross-trigger tolerance criterion (CTTC)
set at 0.3. The Inter-F1 is employed to provide a robust measure of the
overlap between the predicted and actual labels of target sound events, while the
Seg-F1 evaluates the model's performance in accurately identifying sound events
within fixed-length (1s in this study) segments of an audio stream.
These metrics have been widely used for evaluating sound event detection
systems as in \cite{SED1, SED2, NTU}.

In addition to these F1 measures, we also use a false positive event (FPE) count per hour,
similar to the false alarm (FA) count per hour used in keyword spotting (KWS) tasks,
to evaluate the error trigger rate of ICSD systems in long-time background environments
without any target events. In practical applications, a very low FPE count is crucial
to ensure the reliability and efficiency of infant cry and snoring detection systems.
This minimizes the chances of false alarms, thereby improving overall system performance.

\subsection{\textbf{Results}}
\subsubsection{\textbf{Overall Results}}

\begin{table}[!htbp]
\renewcommand\arraystretch{1.3}
\caption{Performance Comparison of Different Models on the Real Test Set.}
\label{Result11}
\vspace{0.2cm}
\centering
\scalebox{0.8}{
\begin{tabular}{l|c|c|c|c}
\toprule
\multirow{2}{*}{\textbf{System}} & \multicolumn{2}{c|}{\textbf{Infant Cry}} & \multicolumn{2}{c}{\textbf{Snoring}}\\
\cline{2-5}
& \textbf{Inter-F1} & \textbf{Seg-F1} & \textbf{Inter-F1} & \textbf{Seg-F1}\\
\hline
MT-CRNN & 0.8203 & 0.8497 & 0.8089 & 0.8529 \\
\hline
CRNN+BEATs & 0.8475 & 0.8990 & \textbf{0.8522} & 0.8763 \\
\hline
Competitive CRNN+BEATs & \textbf{0.8805} & \textbf{0.9174} & 0.8254 & 0.8697 \\
\hline
CRNN+BEATs-14 Class & 0.8534 & 0.8919 & 0.8448 & \textbf{0.8765 }\\
CRNN+BEATs+others & 0.8299 & 0.8732 & 0.8341 & 0.8636 \\
\bottomrule
\end{tabular}}
\end{table}

Tables \ref{Result11} and \ref{Result12} demonstrate the ICSD results on five
different systems for the real and synthetic test sets, respectively. By comparing
the results of the MT-CRNN and CRNN+BEATs systems in Table \ref{Result11},
we observe an absolute performance improvement of around 2\%-5\% in both infant cry
and snoring detection, as measured by both intersection-based F1 and segment-based
F1 metrics. This indicates that introducing additional feature maps
derived from the pre-trained BEATs model provides significant complementary
information to the basic mean-teacher CRNN model.

Furthermore, when comparing the proposed competitive CRNN+BEATs results with those
from the CRNN+BEATs system, we achieve an absolute 3.3\% improvement in
intersection-based F1 for infant cry detection. This suggests that modeling
music, speech and noise together with the target infant cry and snoring as separate
categories can effectively reduce acoustic confusion between infant cry and
background music, speech and noise. However, we observe a slight performance
degradation in snoring detection. This may be due to the model balancing acoustic
discrimination and confusion between noise and snoring. As a result, some
snoring events are misclassified as noise during the inference on the
real snoring test set.

In Table \ref{Result12}, all the results on the synthetic test tests are presented.
Compared to Table \ref{Result11}, all five models perform much better on the synthetic test sets
than on the real ICSD test sets. For example, both the intersection-based F1 and segment-based F1 metrics
of five different models are all above 95\% in both infant cry and snoring detection.
This indicates that the quantity of synthetic strongly labeled data in the train set far exceeds
that of real strongly labeled data, which could lead to potential better models to these synthetic data,
hence resulting in the better performance. Moreover, real strongly labeled data often contain
more complex factors, such as variations in sound quality and event intervals, which may affect
model performance. In contrast, synthetic datasets are usually simpler and clearer,
enabling the models to learn and predict better.

\begin{table}[htbp]
\renewcommand\arraystretch{1.3}
\caption{Performance Comparison on the Synthetic Test Set.}
\label{Result12}
\vspace{0.2cm}
\centering
\scalebox{0.8}{
\begin{tabular}{l|c|c|c|c}
\toprule
\multirow{2}{*}{\textbf{System}} & \multicolumn{2}{c|}{\textbf{Infant Cry}} & \multicolumn{2}{c}{\textbf{Snoring}}\\
\cline{2-5}
& \textbf{Inter-F1} & \textbf{Seg-F1} & \textbf{Inter-F1} & \textbf{Seg-F1}\\
\hline
MT-CRNN & 0.9519 & 0.9527 & 0.9744 & 0.9566 \\
\hline
CRNN+BEATs & \textbf{0.9878} & 0.9769 & 0.9821 & \textbf{0.9722} \\
\hline
Competitive CRNN+BEATs & 0.9845 & 0.9755 & 0.9815 & 0.9672 \\
\hline
CRNN+BEATs-14 Class & 0.9832 & 0.9760 & 0.9850 & 0.9690 \\
CRNN+BEATs+others & \textbf{0.9878} & \textbf{0.9786} & \textbf{0.9901} & 0.9702 \\
\bottomrule
\end{tabular}}
\end{table}

In addition, when comparing the results of the competitive CRNN+BEATs system with
the CRNN+BEATs system, there is almost no performance difference between them on synthetic
test sets. However, on the real infant cry test set, the competitive CRNN+BEATs
significantly outperforms CRNN+BEATs. This indicates that including more categories
(speech, music, noise) helps the model better generalize to real-world scenarios
where such sounds are present. For the real snoring test set, the competitive CRNN+BEATs
shows slightly lower performance than CRNN+BEATs. This could be because the
inclusion of more categories may cause some snoring events to be misclassified as noise.

Furthermore, motivated by the competitive CRNN+BEATs performance, we also performed
the `CRNN+BEATs-14 Class' and `CRNN+BEATs+Others' systems for a fair system comparison
in both Table \ref{Result11} and \ref{Result12}.`CRNN+BEATs-14 Class' means train
a 14-class CRNN+BEATs model using balanced training data of 14 classes
(two target classes: infant cry and snoring, 12 background classes as shown
in Table \ref{tab:sbss}.). The `CRNN+BEATs+Others' model, on the other hand, was
trained as a 3-class system that combines the 12 background classes into a single universal background
class labeled `Others'. Detailed data organization for these two systems can be found on
our ICSD open-source website.

By comparing the results of last three lines in Table \ref{Result11} and \ref{Result12},
we see that on the real infant cry test set, the competitive CRNN+BEATs model shows
the best performance with the highest
Inter-F1 (0.8805) and Segment-F1 (0.9174) scores, suggesting that modeling five categories
enhances the model’s ability to differentiate between these sounds and the infant cry in
real-scenarios. Treating all background sounds as a single category in `CRNN+BEATs+Others'
may not be as effective in distinguishing infant cries from various background noises.
And introducing more background sounds categories as modeling targets in `CRNN+BEATs-14 Class'
may bring more misclassified possibility of target infant cry detection.
On the real snoring test set, these findings differ slightly from those observed
for the infant cry test set. Here, the CRNN+BEATs model achieves the highest F1 scores,
suggesting that it handles snoring detection more effectively when focusing solely
on the two target categories. This may be attributed to the significant acoustic
differences between the two target sounds, infant cries, and snoring.

\begin{table*}[!htbp]
\renewcommand\arraystretch{1.3}
\caption{Training Data ablation experiments of CRNN+BEATs on the Synthetic and Real Test Sets (including
both infant cry and snoring).}
\label{Result22}
\vspace{0.2cm}
\centering
\scalebox{1.0}{
\begin{tabular}{l|l|c|c|c|c}
\toprule
\multirow{2}{*}{\textbf{ID}} & \multirow{2}{*}{\textbf{Training data}} & \multicolumn{2}{c|}{\textbf{Synthetic Test Set}} & \multicolumn{2}{c}{\textbf{Real Test Set}}\\
\cline{3-6}
&& \textbf{Inter-F1} & \textbf{Seg-F1} & \textbf{Inter-F1} & \textbf{Seg-F1}\\
\hline
A0 & Real strongly labeled only & 0.8819 & 0.8835 & 0.8104 & 0.8521 \\
\hline
A1 & Synthetic strongly labeled only & 0.9858 & 0.9773 & 0.7459 & 0.8413 \\
\hline
A2 & Synthetic + Real strongly labeled & \textbf{0.9896} & \textbf{0.9774} & 0.8122 & 0.8635 \\
\hline
A3 & All strongly + weakly labeled & 0.9827 & 0.9724 & \textbf{0.8425} & \textbf{0.8815} \\
\bottomrule
\end{tabular}}
\end{table*}

\subsubsection{\textbf{Ablation Study of Different Training Data Sets}}

Considering the significant role that training data plays in shaping the
performance of a model, Table \ref{Result22} provides an ablation study using
different combinations of training data to observe their effects on system performance.
All the experiments are performed on the CRNN+BEATs system.

\begin{figure}[!htbp]
\caption{False Positive Events per Hour on Non-target Test Set for (a) Infant Cry and (b) Snoring.}
\label{FPIFPS}
\centering
  \includegraphics[width=\linewidth]{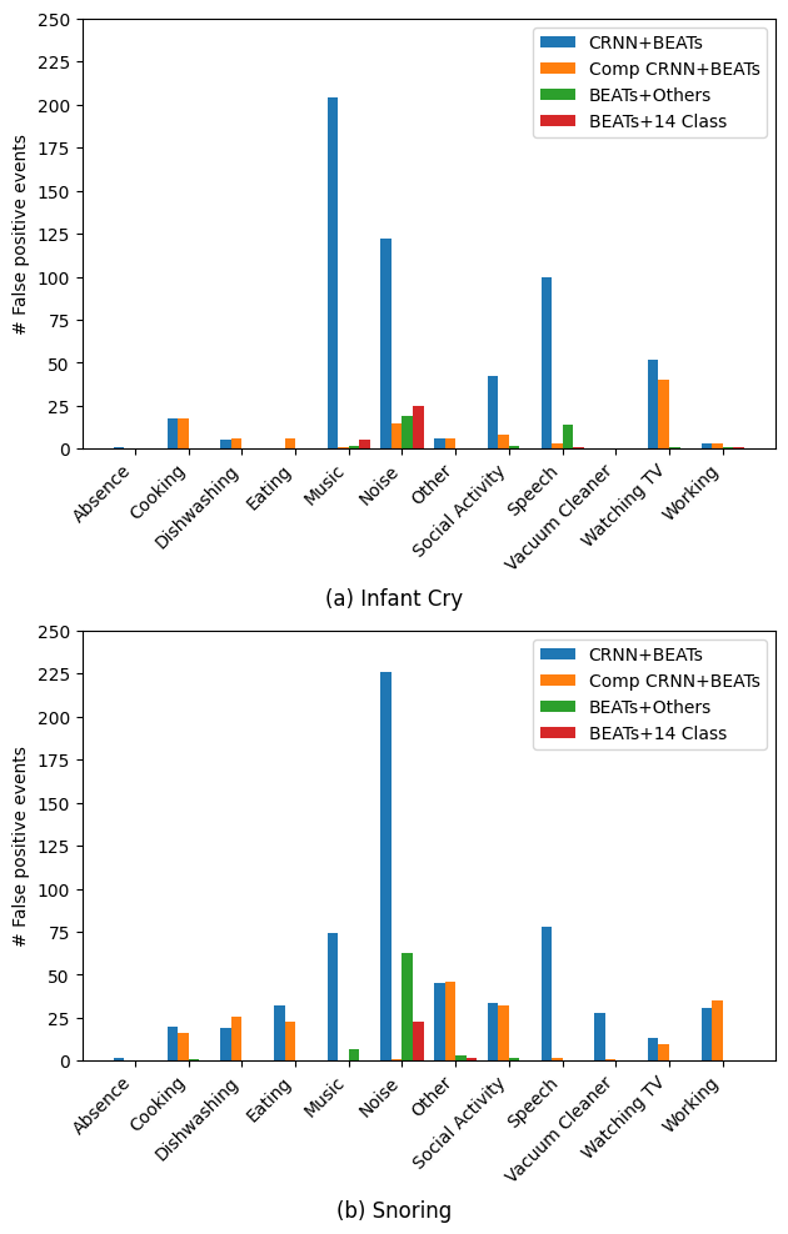}
\end{figure}

In Table \ref{Result22}, we observe a significant performance gap between the synthetic and
real ICSD test sets. The F1 measures on the synthetic set are much better than those on the real set,
indicating that the synthetic data is much easier and cleaner compared to the real test set.
By comparing A0 with A1, we see a large acoustic mismatch between the synthetic and real ICSD data.
In A1, training the model using only the synthetic training set (shown in Table \ref{tab:ovel})
results in high F1 measures on the synthetic test set, while the F1 measures on the real test set
are significantly reduced. In A2, adding the synthetic training data to the real training set improves
performance on both the synthetic and real test sets. This suggests that, despite being simpler,
synthetic data can still provide complementary information to the real training data.
In A3, adding weakly labeled data to the strongly labeled set yields the best results on the real test set,
while maintaining almost the same good results on the synthetic test set as in A2.
This indicates that both weakly labeled data and synthetic training data are important for augmenting
the limited real training data set. Therefore, in all of Table \ref{Result11}, Table \ref{Result12},
and Fig.\ref{FPIFPS}, the models are trained using the entire training set, as shown in Table \ref{tab:ovel},
to include all strongly labeled and weakly labeled data.

\subsection{\textbf{Results on Non-target Test Set}}

As the importance of false alarm in keyword spotting tasks, the false positive rate
on non-target test set is also extremely important for the ICSD tasks in real-world applications.
Because a high false positive rate for target events will greatly
reduce the user's satisfaction in actual product applications. For example,
if an alert is sent to parents every time an infant's cry is detected, a high false
positive event (FPE) rate will end up irritating the users.
Fig.\ref{FPIFPS} presents the FPEs per hour on a `Non-target Test Set' with 12 background classes,
with 1 hour data for each class randomly collected from a total of 232.8 hrs of
background sounds as listed in Table \ref{tab:sbss}. Fig.\ref{FPIFPS} (a) shows
the FPEs for infant cry while (b) shows the results for snoring FPEs.

For the CRNN+BEATs model, the most common false positive detection were music, noise, and speech,
indicating that the model often misclassify these non-target sounds as infant cry or snoring.
In contrast, the Competitive CRNN+BEATs model shows a significant reduction in FPEs for these classes,
indicating the competitive model shows better discrimination ability between target and non-target sounds,
thereby significantly minimizing false positive detection counts in music, noise, and speech conditions.
However, when comparing the FPEs between `Competitive CRNN+BEATs' and the `CRNN+BEATs-14 Class',
`CRNN+BEATs+Others' systems, we find that the `CRNN+BEATs-14 Class' achieves the lowest overall
FPEs of background sounds for both infanct cry and snoring sounds. However,
from Table \ref{Result11} and \ref{Result12},
we see that the `CRNN+BEATs-14 Class' may bring more target sounds misclassifying to other
non-target background sounds at some extent, therefore, there is a performance trade-off between
F1 measures and FPEs in different real application scenarios. We hope these preliminary results
can provide a basic reference for other researchers in the ICSD field.

\section{\textbf{Conclusion}}

In this study, we introduced the ICSD dataset, a new publicly available resource
designed to bridge the gap in the availability of high-quality, strongly and weakly
labeled data for the detection of infant cries and snoring sounds. Besides the
open-source dataset, we also provided several baseline ICSD systems to show
extensive preliminary ICSD results on this dataset as reference points for
future ICSD studies, including a mean-teacher based CRNN, a CRNN+BEATs,
and competitive CRNN+BEATs systems. Based on these baseline systems,
we performed detailed analysis on both the real and synthetic
test sets, as well as on a purely non-target background test set to evaluate the
degree of false positive rates in different types of background
acoustic environments. We hope that our ICSD dataset will be beneficial for the advancement
of sound event detection in home environments, and all these preliminary experimental findings and observations
on this released dataset can serve as a solid foundation for future innovations in this area.
In addition, it is worth noting that the current usage of our released synthetic and background
datasets provided in the ICSD is just a basic example. Researchers can re-organize and use our data simulation
script to synthesize any size/any types of synthetic dataset for their ICSD system building.
Our future work will focus on exploring new methods to improve ICSD performance and
continuously enhancing this dataset to achieve larger and wider domain coverage.

\section*{Acknowledgments}
The work is supported by the National Natural Science Foundation of China (Grant No.62071302). The authors would like to thank Dr. Carlos Alberto Reyes Garcia for providing us the Baby chillanto database.

\bibliography{ICSD}

\end{document}